# INVESTIGATION OF THE HYDRO-METEOROLOGICAL HAZARDS ALONG THE BULGARIAN COAST OF THE BLACK SEA BY RECONSTRUCTIONS OF HISTORICAL STORMS

V. GALABOV[*], A. KORTCHEVA, A. BOGATCHEV, B. TSENOVA

*National Institute of Meteorology and Hydrology, Bulgarian Academy of Sciences (NIMH-BAS), 66Tsarigrasko Shose, 1784 Sofia, Bulgaria*

*E-mail*: *vasko.galabov@meteo.bg*

**Abstract**. Information about the hydro-meteorological parameters during the extreme sea storms is of significant importance for the sustainable development in the context of flood risk for the coastal areas. Usually there is a lack of sufficiently long history of instrumental measurements of the extreme winds, waves and storm surges. Simulation of historical storms is an important tool to evaluate the potential coastal hazards. In the absence of measured data hindcasts can satisfy the need for historical data. The wave and storm-surge regional numerical simulations have been carried out for the ten most severe storms over the Bulgarian coast of the Black Sea from the period 1972-2012. The ERA-Interim and ERA-40 reanalysis of wind at 10 m and mean sea level pressure have been downscaled with a high resolution atmospheric model ALADIN to the horizontal and time scales suitable for precise evaluation of hydro-meteorological parameters during the storms. The downscaled fields of wind and sea level pressure have been used as input for the wave and storm surge models.

*Keywords*: Black Sea, waves, storm surge, coastal hazards, storms.

---

[*] For correspondence.

AIMS AND BACKGROUND

The strong winds, the high waves and sea level increase due to storm surges during the sea storms are major hazards for the coastal areas, causing coastal flooding, damage to the infrastructure, coastal morphodynamic changes and possible loss of human life. The Black Sea storms are most severe at the Western part and the North-Eastern part of the Black Sea[1-3]. Some notable cases are the storm of February 1979 that caused massive damages of the Bulgarian coast and flooding of urban areas, and the recent severe storm of February 2012. Among the important studies of the Black Sea storminess are the articles of Valchev et al..[4], Arkhipkin et al..[1], Surkova et al..[2], Polonsky et al..[4], Rusu et al..[5]. Valchev et al..[4] find out that there is no significant trend of the storminess but there is a period with increased storminess during the sixties and the seventies. Arkhipkin et al..[1] confirms these findings and argue that the long term changes are associated with negative values of the North Atlantic Oscillation (NAO) anomaly. Link between NAO and the Romanian coastal climate was also found, but for the rainfall regime by Bandoc et al..[6]. In the study of Polonsky et al..[4], the authors present evidence for a low frequency oscillation of the Northern Black Sea storminess with a period of 50-70 years, but instead of the weak dependence on the NAO they present an evidence that it is associated with the variation of the Atlantic Multidecadal Oscillation (AMO) and it is interplay with the Pacific Decadal Oscillation (PDO). An important finding in that study is that any attempt to obtain the return periods of the extreme waves based on measurement data for the recent 20-30 years or too short wave hindcasts will result in an underestimation of the probability of extreme events due to the significantly higher probability of occurrence of events during the periods with negative AMO and PDO anomaly. This finding questions the validity of the estimations of return periods for the Black Sea that do not include the period prior to 1980. In their study Surkova et al.[2] are using climate projections to investigate the future Black Sea storminess and conclude that a significant decrease of the frequency of storms is expected due to the climate change. As it is shown in the recent study of Kravtsov et al.[7] the models reproduce multidecadal climate variations that are weaker than the real variations which means that a future decrease of the Black Sea storminess is not in a contradiction with a possible increase during the next few decades. Due to the lack of instrumental measurements of the wave parameters in the Black Sea, the only approach to obtain the return periods of the extreme events is to perform wave and storm surge hindcasts. Hindcasts may be based on the usage of atmospheric reanalysis data. The most widely used reanalysis such as ERA-Interim[8] and NCEP Reanalysis II[9] start from 1979. The period before that is known as a period of elevated storm frequency for the Black Sea. Moreover, the reanalyses are with low spatial and temporal resolution and the wind fields deviate significantly from the observations. ERA-Interim underestimates while NCEP Reanalysis II overestimates the wind speeds (Kara et al.[10]). Our approach is based on the reconstructions of historical storms by the usage of dynamical downscaling of the ERA-Interim and ERA40 atmospheric fields with a high resolution regional atmospheric model. The downscaled winds at 10 m height and mean sea level pressure fields (MSLP) are used as an input to a spectral wave model and a storm surge model in order to reconstruct the selected events[11]. We present our modelling approach and the reconstructions of the two most significant storms from a set of selected 10 historical storms.

EXPERIMENTAL

The dynamical downscaling procedure is implemented to reproduce the atmospheric fields over the Black Sea with a high resolution. Reanalysis of the atmospheric fields of ERA-Interim[8] for the storms after 1978 and ERA40 with coarse spatial (80 and 125 km, respectively) and temporal (6 h) resolution is dynamically downscaled to a resolution of 10km and 1 h temporal resolution. The downscaling is performed by the usage of the limited area atmospheric model ALADIN[12]. ALADIN is a spectral model for regional meteorological forecasts, developed by an international cooperation led by METEO-FRANCE. The initial and lateral boundary conditions for the model runs are obtained from the transformed reanalysis data in ARPEGE global atmospheric model format provided by METEO-FRANCE (Bresson et al.[13]). The wave model used in this study is SWAN (Simulating WAves Nearshore) version 40.91ABC (Booij et al.[14]). SWAN is a third generation wave model that is designed to simulate the waves in the near shore waters and is applied often in semi enclosed seas, estuaries and lakes. SWAN simulates the wave generation, dissipation, nonlinear interactions between the waves by solving the spectral wave action balance equation[14]. The model was often implemented and studied for the Black Sea applications (Butunoiu and Rusu[15], Ivan et al.[16], Rusu et al.[17], Akpinar et al.[18, 19], Fomin et al.[20]). The computational grid, that we use, is a spherical regular grid with 0.0333° horizontal resolution and covers the entire Black Sea. Different available source terms parameterisations are tested in order to choose the most suitable one. The spectral discretisation is based on 36 directions and 30 frequencies logarithmically spaced from 0.05 to 1.00 Hz. The simulations of the storm surges are based on the depth integrated storm surge model of METEO- FRANCE[21] adapted to the Black Sea by Mungov and Daniel[11]. The model grid is identical to the model grid of SWAN. The values of the bottom friction are 0.0015 over the shelf and 0.000015 over the liquid bottom that was set at the level of the Black Sea mixed layer depth. Data for seasonal variations of the Black Sea mixed layer depth are taken from the study of Kara et al.[22]. Data used for the validation of the wind fields are from the ASCAT scaterometer on board of METOP-A satellite. Data used for the validation of the wave model are along the track satellite altimeter data from the JASON and ENVISAT satellites. The *in-situ* wave data for the storm of February 2012, measured by acoustic Doppler current profiler (ADCP) were provided by the Bulgarian Oceanological Institute (Trifonova et al.[23] and Valchev et al.[24]).

RESULTS AND DISCUSSION

The storm of 7-8 February 2012 is the most severe storm[25] since 1979 that caused coastal flooding and significant damages along the Bulgarian coast. Chervenkov[26] found out that the cyclone is the second most intensive after 1947. The downscaled 10 m wind fields are validated by a comparison with remotely sensed wind speeds. The statistics are shown in Table 1. The bias is below 5% of the mean measured wind speed and the scatter index (SI) below 10%. The downscaling of the atmospheric fields can be considered successful with a small underestimation that cannot influence significantly the marine models. The simulations of the significant wave height (hereafter denoted SWH) are shown in Fig. 1. The usage of ERA-Interim data leads to SWH 0.4-0.8 m lower than the results when downscaling is used. The model output is compared with the satellite measurements along the tracks of JASON and ENVISAT satellites that crossed the Western Black Sea during the storm. The comparison statistics are shown in Table 2. The statistics for all 4 tracks are presented (214 measurements) together with a separate statistics for one track during the peak phase of the storm (54 measurements). We tested three different parameterisations of the wave energy generation by wind and wave energy dissipation due to wave breaking in deep water (whitecapping)- the parameterisations of Komen[27], Janssen[28] and Westhuysen[29] (wind input parameterisation of Janssen combined with the saturation based parameterisation of Westhuysen). The Komen source terms overestimates in the case of the high resolution wind input while all other choices underestimate slightly the SWH. The lowest biases are obtained using the Westhuysen source terms (and the bias does not increase during the peak of the storm) as well as the SI for the track corresponding to the peak of the storm. For all tracks the scatter index when using Janssen source terms is the lowest.

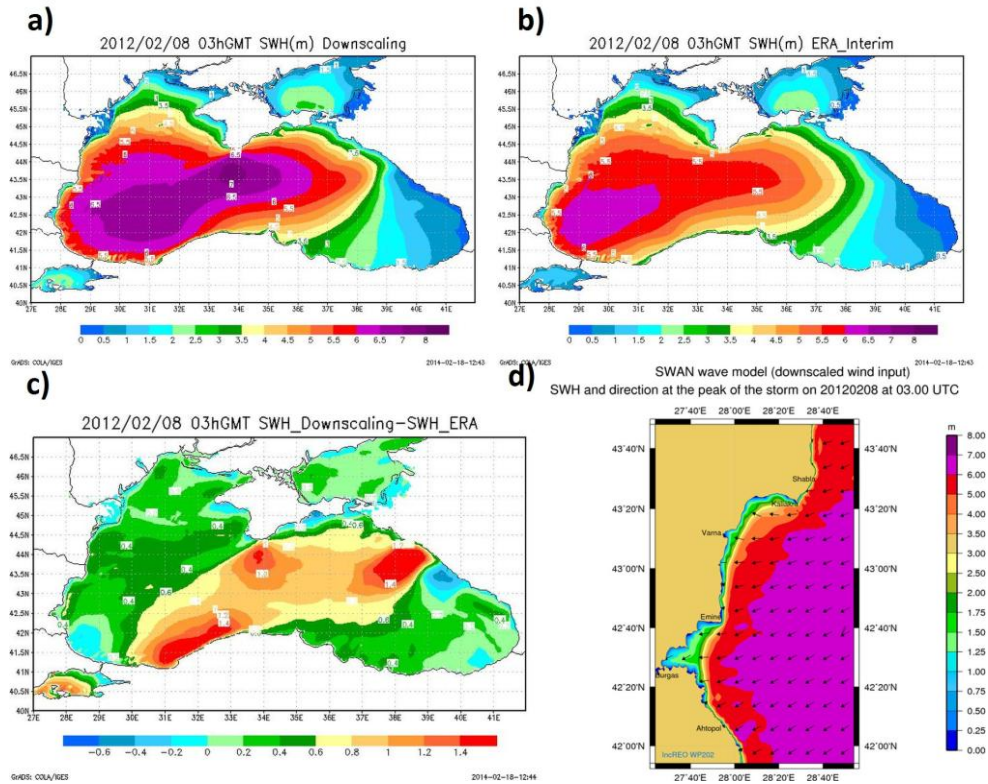

**Fig. 1**. Simulation of the significant wave height fields for the peak storm of 7-8 February 2012 during of first hours of 8 February: *a* - simulation using the downscaled wind fields; *b* - simulations using the coarse resolution ERA-Interim input data; *c* - the difference between the simulated significant wave heights when using the downscaled and reanalysis input data; *d* - simulation of the significant wave height for the Bulgarian coast

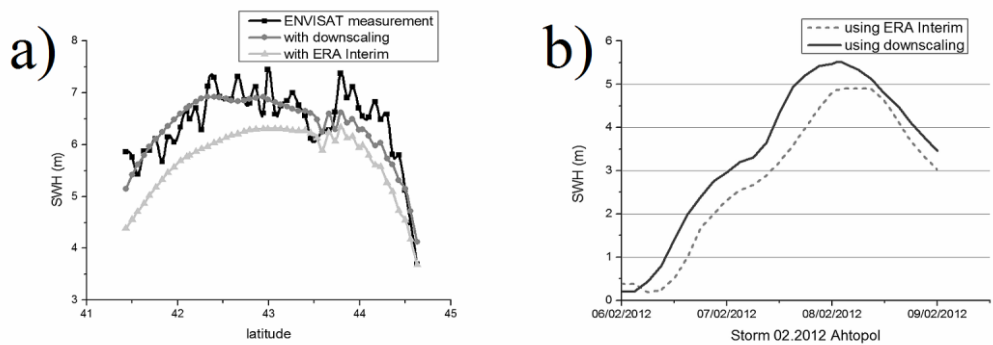

**Fig. 2.** Comparison of the simulated significant wave height using the downscaled input data and ERA-Interim and the measurements along the satellite track of ENVISAT satellite in the evening of 7 February (*a*); comparison of the time series of the significant wave height for Ahtopol with the high and low resolution wind input (the southernmost Bulgarian coast which was the most affected by the storm) (*b*)

**Table 1.** Statistics of comparison the downscaled wind speed and ASCAT data (7-8.02.2014)

| Observation mean (m/s) | Model mean (m/s) | Bias (m/s) | RMSE (m/s) | Scatter index |
|---|---|---|---|---|
| 13.80 | 13.38 | -0.43 | 1.26 | 0.09 |

**Table 2.** Statistics of the comparison of SWAN model simulated significant wave heights and remotely sensed significant wave height. In each cell the upper values is for all tracks during the storm and the lower value for the track that crossed the Western Black Sea during the peak of the storm

| Type of wind input | Type of parameterisation scheme | Observation mean (m) | Model mean (m) | Bias (m) | Root mean square error (m) | Scatter Index |
|---|---|---|---|---|---|---|
| Downscaled | Komen | 4.73 | 4.90 | +0.17 | 0.67 | 0.14 |
|  |  | 6.38 | 6.69 | +0.30 | 0.39 | 0.08 |
| Downscaled | Janssen | 4.73 | 4.59 | -0.14 | 0.58 | 0.12 |
|  |  | 6.38 | 6.14 | -0.24 | 0.50 | 0.08 |
| Downscaled | Westhuysen | 4.73 | 4.62 | -0.08 | 0.68 | 0.14 |
|  |  | 6.38 | 6.30 | -0.08 | 0.37 | 0.06 |
| ERA Interim | Komen | 4.73 | 4.20 | -0.53 | 0.76 | 0.16 |
|  |  | 6.38 | 5.68 | -0.70 | 0.78 | 0.12 |
| ERA Interim | Westhuysen | 4.73 | 3.92 | -0.81 | 0.99 | 0.21 |
|  |  | 6.38 | 5.30 | -1.08 | 1.14 | 0.18 |
| ERA Interim | Janssen | 4.73 | 3.84 | -0.89 | 1.03 | 0.22 |
|  |  | 6.38 | 5.20 | -1.18 | 1.24 | 0.19 |

However, taking into account the higher negative bias in this case, we may conclude that the parameterisation scheme of Westhuysen leads to slightly better results in comparison to other schemes. Figure 2*a* presents the comparison of the simulated SWH using the two wind inputs and the Westhuysen physics with the measurements of ENVISAT during the maximum phase of the storm. The advantage of the usage of high resolution downscaled winds is obvious. Simulations of the waves are also compared with in-situ measurements during the storm at 20m depth at Pasha Dere beach.

The comparison of the maximum measured and modelled values at that location is presented in Table 3. The use of Westhuysen and Janssen source terms results in values of maximum SWH that correspond very well with the measurements and the usage of Komen physics results in slight overestimation of the SWH. The use of ERA-Interim wind input leads to significant underestimation of SWH. The simulation of the storm surge during the event with the downscaled and ERA- Interim wind forcing resulted in values of the maximum surge (Fig. 3) that are higher by more than 30% when using the downscaled atmospheric fields.

**Table. 3**. Comparison of the measured by ADCP maximum significant wave height during the storm of 2012 and the modelled with different choices of parameterisations and with the ERA-Interim wind input

|  | ADCP measurement | SWAN, Komen | SWAN, Westhuysen | SWAN, Janssen | SWAN Komen, ERA-INTERIM input |
|---|---|---|---|---|---|
| Significant wave height (m) | 4.77 | 4.99 | 4.67 | 4.89 | 4.1 |
| Peak wave period (s) | 11.1 | 12.3 | 12.0 | 12.4 | 11.5 |
| Peak wave direction | 84° | 85° | 85° | 85° | 87° |

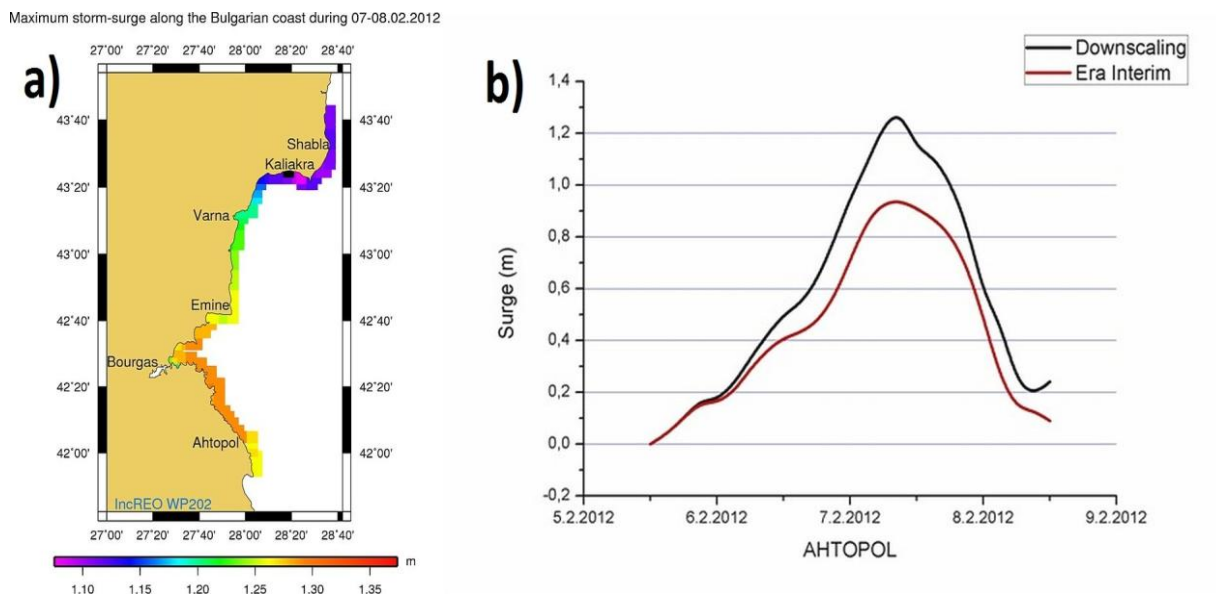

**Fig. 3**. The maximum storm surge along the Bulgarian coast during the storm of February 2012 (*a*) and time series of storm surge at Ahtopol with downscaled and ERA-Interim wind forcing (*b*)

The next storm reconstruction is the storm of February 1979. The storm lasted for more than 10 days with a peak on 19 February. It is notable due to the highest storm surge ever recorded for the Bulgarian coast- the sea level rise was 1.43 m at Irakli tide gauge (Emine cape location) and above 1.5 m at Varna and Bourgas tide gauges. The significant wave height according to Belberov et al.[30] reached 5.8 m at 15 m water depth at the location of the scientific pier at Shkorpilovtsi beach. It is possible to classify this storm as an expected very extreme event during the period of elevated Black Sea storminess during the seventies and possible future period. The reconstruction of the SWH is shown in Fig. 4.

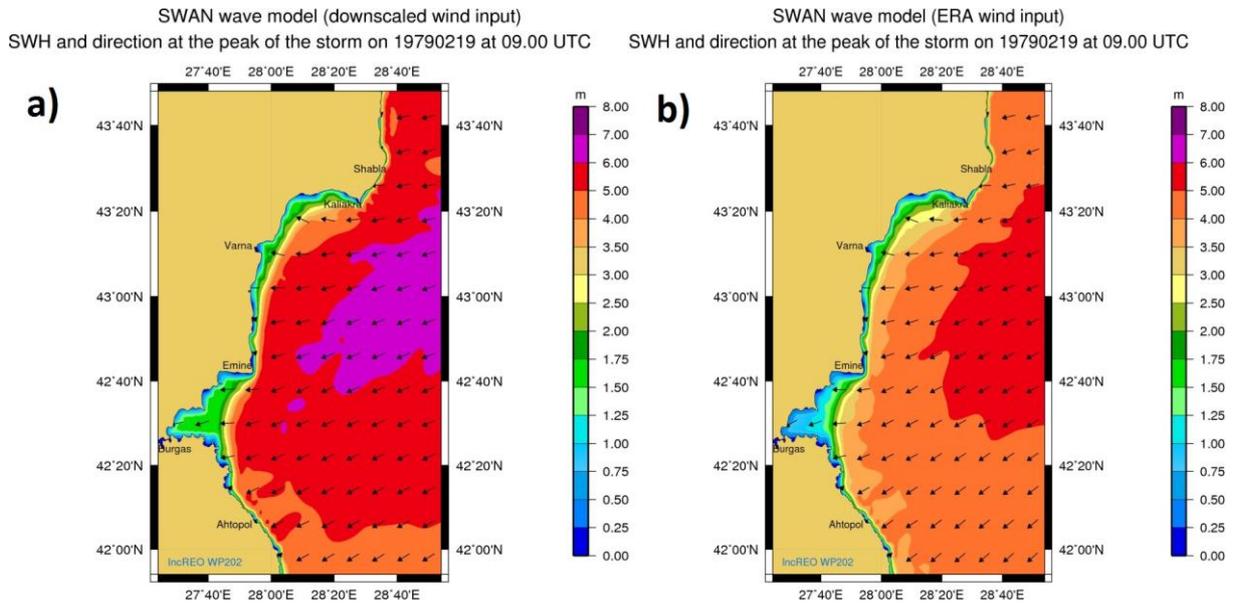

**Fig. 4**. Reconstruction of the significant wave height fields for the Bulgarian Black Sea coast using the downscaled wind input (*a*) and ERA-Interim (*b*)

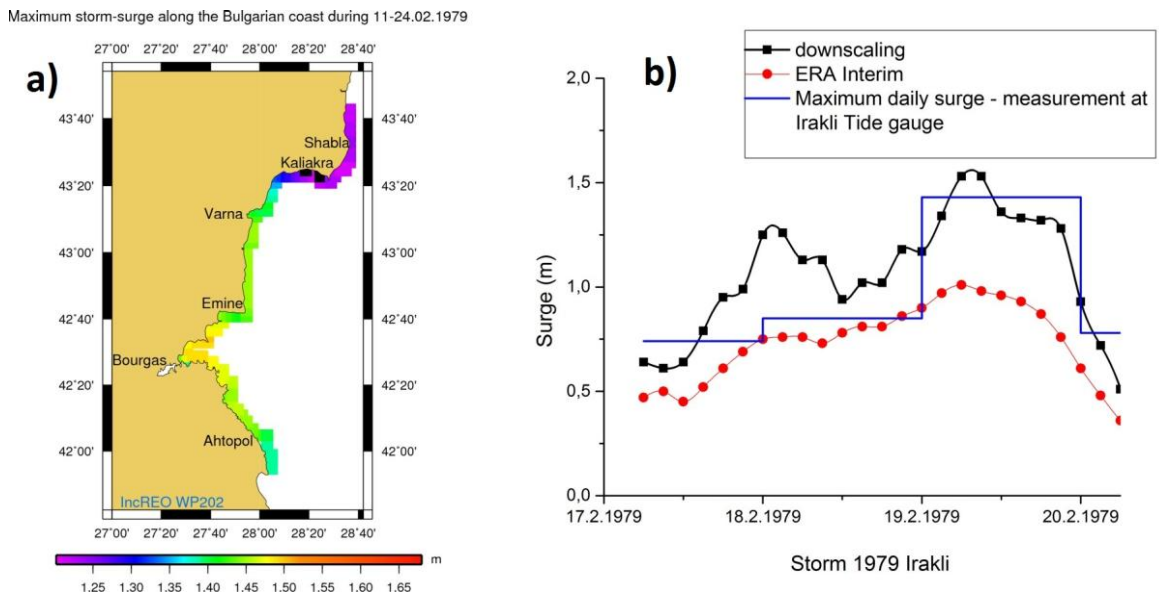

**Fig. 5**. Maximum storm surge along the Bulgarian coast during the storm of February1979 (*a*) and comparison of the simulated storm surge with the daily maximum surge measured at the Irakli tide gauge (*b*)

At the location of Shkorpilovtsi at 15 m depth the estimation with SWAN of the maximum SWH is 5.5 m using the downscaled wind input and 4 m using the ERA-Interim winds. The usage of reanalysis failed to reproduce the intensity of the event. The values of the maximum storm surge during the storm of February 1979 fit well with the available information about the event (Fig.5*a*). Comparison between the measured daily maximum sea levels by the Irakli tide gauge and modelled ones is presented in Fig. 5*b*). The simulation with the downscaled atmospheric fields overestimates the measured maximum of 1.43 m with 9 cm. One way to minimise the bias is to replace the bulk formula for the wind drag coefficient in the storm surge model with a formula that use information of the sea state from SWAN in the form of wave steepness or friction velocity (explicitly computed by SWAN when using the Janssen[28] parameterisation of the wave growth in the wave model). By default the storm surge model is using a linear dependency of the wind drag coefficient on the wind speed in the form of Smith and Banke[31]. We used several alternative formulations of the wind drag coefficient in the storm surge model. We introduced a change in the storm surge model so that the model reads the friction velocity or wave steepness from SWAN. The relation of the wind drag coefficient on the wave steepness and the wind speed is based on the work of Guan and Xie[32]. The wind drag coefficient is defined as:

$$C_d = 0.001*(0.78+K*\delta*U_{10}) \qquad 1)$$

where $C_d$ is the wind drag coefficient; $\delta$ - the wave steepness; $U_{10}$ - the 10 m wind speed, and *K* - a tunnable coefficient. This is a simplified version of the formulation of Guan and Xie and we found that for the Black Sea applications the best results are obtained by using *K*=0.7 (Fig. 6).

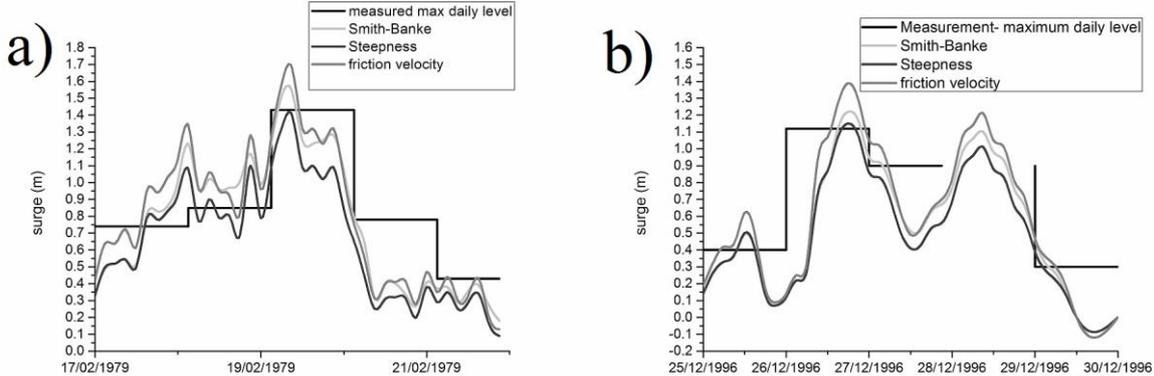

**Fig. 6**. Comparison of the measured daily maximum storm surges with the storm surge model output using various definitions of the wind drag coefficient: *a* - for the storm of 1979; *b* - the same comparison for a storm during December 1996

CONCLUSIONS

Ten historical storm situations in the Western Black Sea have been reconstructed using a high resolution downscaling of the ERA-Interim and ERA40 reanalysis with the limited area atmospheric model ALADIN. The downscaled wind fields have been validated using satellite data seem to be reliable. Simulations of the waves and storm surges have been carried out by the use of SWAN wave model and a storm surge model. The reconstructions of the two most significant storms have been presented. The results from wave simulations have been compared with satellite data and *in-situ* measurements. Different parameterisations of the source terms in SWAN have been evaluated. The storm surge simulations have been compared with the available data from tide gauges. An approach to use the sea state information within the storm surge model has been evaluated. It has been argued that the storm of February 1979 is representative for the expected extreme scenario under future period of increase in the storminess.

**Acknowledgements**. The research was carried out as a part of the IncREO project funded by the European Community Seventh Framework Program under grant agreement Number 312461. The research was partially co-financed by the Ministry of the Education and Science of Republic of Bulgaria.

APPENDIX:

The locations mentioned within the paper:

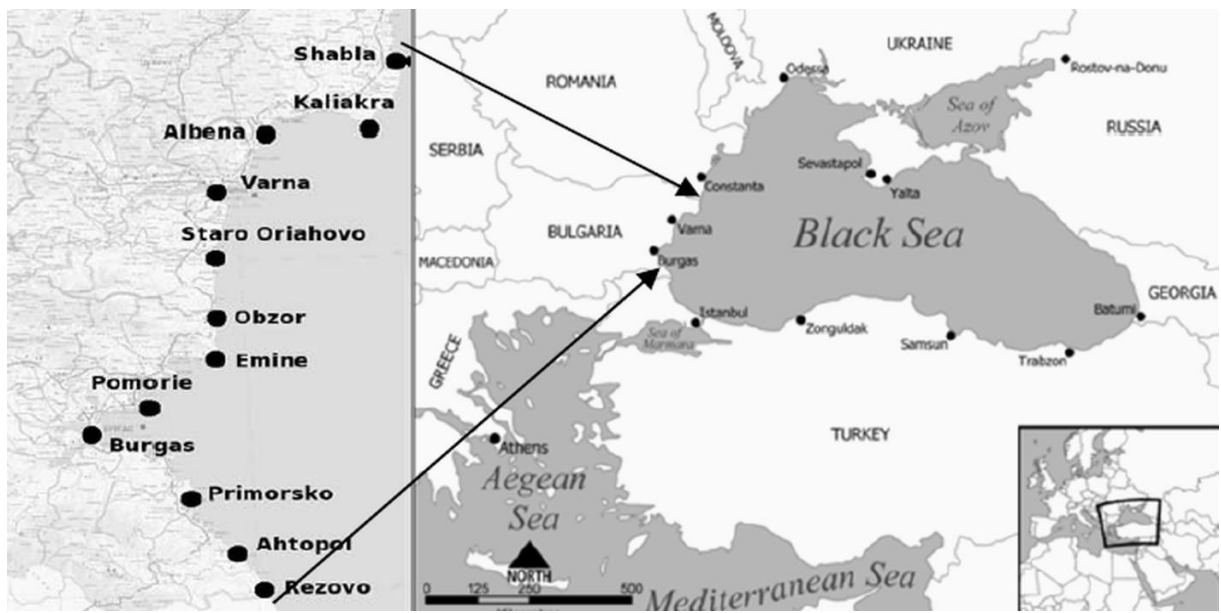